\documentstyle[11pt,aaspp4]{article}

\def\iras{{\sl IRAS}}
\def\etal{{\it et~al.}}
\def\vev#1{\left\langle #1\right\rangle}
\def\kms{\ifmmode {{\rm \ km \ s^{-1}}}
\else{$\rm {km \ s^{-1}}$}\fi}
\def\mpc{\ifmmode {\,h^{-1}{\rm Mpc}}
\else{$\,h^{-1}{\rm Mpc}$}\fi}
\newcommand{\gtsima}{$\; \buildrel > \over \sim \;$}
\newcommand{\ltsima}{$\; \buildrel < \over \sim \;$}
\newcommand{\simgt}{\lower.5ex\hbox{\gtsima}}
\newcommand{\simlt}{\lower.5ex\hbox{\ltsima}}

\lefthead{Strauss et al.}
\righthead{Velocity Dispersion as a Function of Density}

\begin{document}

\title{The Galaxy Pairwise Velocity Dispersion as a
Function of Local Density}

\author{Michael A. Strauss\altaffilmark{1}, Jeremiah
P. Ostriker, and Renyue Cen}
\authoremail{(strauss,jpo,cen)@astro.princeton.edu}
\affil{Dept.~of Astrophysical Sciences, Princeton
University, Princeton, NJ 08544}

\altaffiltext{1}{Alfred P. Sloan Foundation Fellow.}

\begin{abstract}
The standard method of measuring the galaxy pairwise velocity
dispersion on small scales via the anisotropy in the two-point
correlation function in redshift space suffers from the fact that it
is a {\em pair-weighted\/} statistic, and thus is heavily weighted by
the densest regions in a way that is difficult to calibrate.  We
propose a new statistic, the redshift difference of close projected
pairs of galaxies as a function of local density, which is designed to
measure the small-scale velocity dispersion as an explicit function of
density.  Computing this statistic for a volume-limited subsample of
the Optical Redshift Survey, we find that the small-scale velocity
dispersion rises from 220 \kms\ in the lowest density bins to 760
\kms\ at high density.  We calculate this statistic for a series of
mock catalogs drawn from a hydrodynamic simulation of an $\Omega h =
0.5$ Cold Dark Matter universe (standard CDM), and find that the
observed velocity distribution lies $\simgt 1\,\sigma$ below the
simulations in each of eight density bins, formally ruling out this
model at the $7.4\,\sigma$ level, quantifying the
well-known problem that this model produces too high a velocity
dispersion.  This comparison is insensitive to the normalization of
the power spectrum, although it is quite sensitive to the density and
velocity bias of galaxies relative to dark matter on small scales.
\end{abstract}

\keywords{}

\section{Introduction}
\label{sec:introduction}

The relative pair-wise velocity dispersion of galaxies $\sigma_{12}$ on small
($\sim 1 \mpc$) scales has long been used as a diagnostic of
cosmological models.  Peebles (1976a,b) used the assumption of
hydrodynamic equilibrium of the galaxy population on small scales to
derive the Cosmic Virial Theorem (CVT), which relates $\sigma_{12}$ to the
two- and three-point correlation function, and the value of the
Cosmological Density Parameter $\Omega_0$.  However, a number of
authors have questioned the basic assumptions on which the CVT is based
(Fisher \etal\ 1994, hereafter F94; Bartlett \& Blanchard 1996; Suto \& Jing 
1997).   

Davis \etal\ (1985) recognized $\sigma_{12}$ as an important statistic
to compare with cosmological models; indeed, the discrepancy between
the predictions of the standard Cold Dark Matter model and the
observed value of $\sigma_{12}$ led them to adopt the then-new idea
that the galaxy population is strongly biased relative to the
distribution of dark matter (Bardeen \etal\ 1986).  The use of $\sigma_{12}$
as a distinguisher of models has been controversial ever since, and
the question of the relationship between the normalization of the
standard Cold Dark Matter (CDM) model and the resulting $\sigma_{12}$
has been hotly debated in the literature (Ostriker \& Suto 1990; Cen
\& Ostriker 1992; Couchman \& Carlberg 1992; Gelb \& Bertschinger
1994; Zurek \etal\ 1994; Brainerd \& Villumsen 1994; Weinberg 1995;
Brainerd \etal\ 1996; Somerville, Primack, \& Nolthenius 1997).

Davis, Geller, \& Huchra (1978), Peebles (1980), Davis
\& Peebles (1983), and Bean \etal\ (1983) 
introduced what is now the standard method of measuring $\sigma_{12}$.
From a redshift survey of galaxies, one can measure the correlation
function of galaxies in redshift space $\xi(s)$.  Recognizing that peculiar
velocities systematically distort the separation of pairs of galaxies
along the line of sight, one can calculate $\xi$ as a function of
the component of the separation vector both parallel ($\pi$) and
perpendicular ($r_p$) to the line of sight.  In real space, the
lack of a preferred direction means that $\xi(r_p,\pi)$ should be
isotropic, but in redshift space, the
correlations will be elongated along the $\pi$ direction on small
scales, because of the small-scale pairwise motions of galaxies.  This
effect can be modeled as a convolution of the real-space correlation
of galaxies (which can be determined by a projection of $\xi(r_p,\pi)$
onto the $r_p$ axis) with the distribution function of pairwise
peculiar velocity differences, thus allowing a determination of at
least the second moment of this distribution function (F94; Strauss \&
Willick 1995; Marzke \etal\ 1995). Recent 
determinations of $\sigma_{12}$ from 
redshift survey data include F94, Marzke \etal\
(1995), and Guzzo \etal\ (1996, 1997). 

However, the determination of $\sigma_{12}$ by this method is quite
unstable.  Because it is based on the two-point correlation function,
$\sigma_{12}$ is pair weighted, and thus is heavily weighted by the
densest regions of a sample.  Because these regions naturally have the
highest velocity dispersion (as one can show analytically with a
straightforward extension of the CVT; Kepner, Summers, \& Strauss
1997, and as we will show explicitly below), this statistic is
strongly dependent on the presence or absence of rare, rich clusters
within a sample (Mo, Jing, \& B\"orner 1993; Marzke \etal\ 1995;
Somerville, Davis, \& Primack 1997; Guzzo \etal\ 1996, 1997).
Moreover, it has been recognized for at least a decade that outside of
clusters, the velocity field is very cold (Brown \& Peebles 1986;
Sandage 1986; Burstein 1990; Groth, Juszkiewicz, \& Ostriker 1989;
Ostriker \& Suto 1990; Strauss, Cen, \& Ostriker 1993; Willick \etal\
1997) and it is hoped that a direct measurement of the velocity
dispersion in the field would yield an even stronger constraint on
cosmological models than does the global value of $\sigma_{12}$.

Indeed, the sensitivity of $\sigma_{12}$ to high density, high
velocity dispersion regions in both observations and simulations is
the cause of much of the controversy over the comparison between
models and real data referred to above.  These high-density regions
are intrinsically rare, and thus small observed or simulated volumes
will not contain any high velocity dispersion virialized structures.
Cen \& Ostriker (1994) found that the rms one-dimensional velocity
dispersion for particles at 1\mpc\ separation in a Mixed Dark Matter
simulation increased from $\sim 400 \kms$ to $\sim 600 \kms$ as the
box size was increased from 25\mpc\ to 320\mpc, at which point it
finally converged.  On the observational side, this effect produces a
large cosmic variance in the $\sigma_{12}$ statistic (cf., Marzke
\etal\ 1995).  To combat this effect, some workers have omitted high
density regions both from observed and simulated samples, but the
results are quite sensitive to exactly how such regions are excised,
and of course one is is potentially throwing out valuable information
in doing so.

Cen \& Ostriker (1994; cf., Fig.~18b) showed from their Mixed Dark
Matter simulation that the pairwise velocity dispersion is a strong
function of density, smoothed on a scale of 5-10\mpc.  This motivates
us to develop a new measure of the pairwise velocity dispersion of
galaxies from a redshift survey, {\em as a function of local
density\/} (for related approaches, see Kepner \etal\ 1997; Davis,
Miller, \& White 1997).  Although the statistic we define here is not
mathematically identical to $\sigma_{12}$ as conventionally defined,
we show that it is a reasonable approximation.  Moreover, we can
calculate this statistic from Monte-Carlo realizations of our
observations, drawn from simulations of various cosmological models.
We can therefore use observations to compare with models, and rule
them out, as the case may be.

Our primary purpose in this paper is to
present our technique of measuring the velocity dispersion as a
function of density.  We calculate this statistic from available data,
and make a preliminary comparison with the best available
simulations. 
We introduce our statistic in \S~\ref{sec:method}, and present results
using the {\it Optical Redshift Survey} (ORS) of Santiago \etal\
(1995, 1996; hereafter S95, S96).  In \S~\ref{sec:nbody}, we compare
our results to those found by applying our technique to Monte-Carlo
simulations drawn from hydrodynamic and $N$-body simulations of
various models of structure formation.  Our conclusions may be found
in \S~\ref{sec:conclusions}.  

\section{The Pairwise Velocity Dispersion as a function of Local
Density}
\label{sec:method}

Galaxies show strong clustering on small scales.  Therefore a pair of
galaxies whose separation $r_p$ on the plane of the sky is small is
likely to have a small separation in real space.  If the peculiar
velocity difference of the pair is large relative to the real space
separation expressed in velocity units, then the separation $\pi$
along the line of sight is largely a measure of this peculiar velocity
difference.  Motivated by this simple mental picture, we will examine
all pairs of galaxies whose $r_p$ (as defined by F94) is smaller than
1 \mpc.  We then look at the distribution of $\pi$, the redshift space
distance between these.  This distribution is closely related to the
correlation function along the line of sight $\int_0^1 d r_p
\xi(r_p,\pi)$; indeed, it differs from this quantity only in not being
normalized by the distribution expected in an unclustered universe for
the given survey geometry.

We can define a local density associated with each galaxy (in redshift
space) by smoothing the galaxy distribution with a Gaussian of
standard deviation 400 \kms.  We then simply define the density
associated with any pair of galaxies as the average of the densities
associated with each one individually.  The 400 \kms\ smoothing scale
is chosen to be appreciably higher than the scale on which we examine
the small-scale velocity dispersion, and large enough so that on
average, several galaxies are included within the smoothing window,
yet small enough that we can still refer to a {\em local\/} density
field.

The following are our considerations of the appropriate data to use
for this analysis:
\begin{enumerate}
\item The sample should be {\em well-defined}, to allow the local
density to be determined, and to allow comparisons with results from
simulations; 
\item The sample should be {\em volume-limited}, so that the
distribution of redshift differences not be a function of
distance from the observer; 
\item The sample should be {\em shallow}, to minimize the
number of chance projections on the sky;
\label{item:shallow}
\item The sample should have {\em large solid angle}, to maximize the
survey volume given the constraint of item~\ref{item:shallow};
\item The sample should be {\em dense}, to maximize the number of pairs
at small separation.
\end{enumerate}
The survey at our disposal that best fits these criteria is the ORS
(S95, S96), and in particular, the $m = 14.5$-limited
subsample of it (ORS-m in the notation of S95).  The sample
covers 6.62 ster, consists of 5697 galaxies, and is drawn from three
distinct galaxy catalogs in different regions of the sky.  For each
galaxy, we define a local density contrast $\delta$ following the
techniques of S96.  All calculations are done in
redshift space, as measured in the rest frame of the Local Group, but
for the density determination, we follow S96 in collapsing galaxies
associated with several nearby clusters to a common
redshift\footnote{This is of course not done in computing our redshift
difference statistic!} (cf.,
Yahil \etal\ 1991).  We then draw a volume-limited subsample to 3000
\kms, leaving 1123 galaxies.  From this subsample, we identify all pairs
with $r_p < 1 \mpc$, and measure $\pi$, the radial component of the
redshift difference vector between them. 

\begin{figure}
\plotone{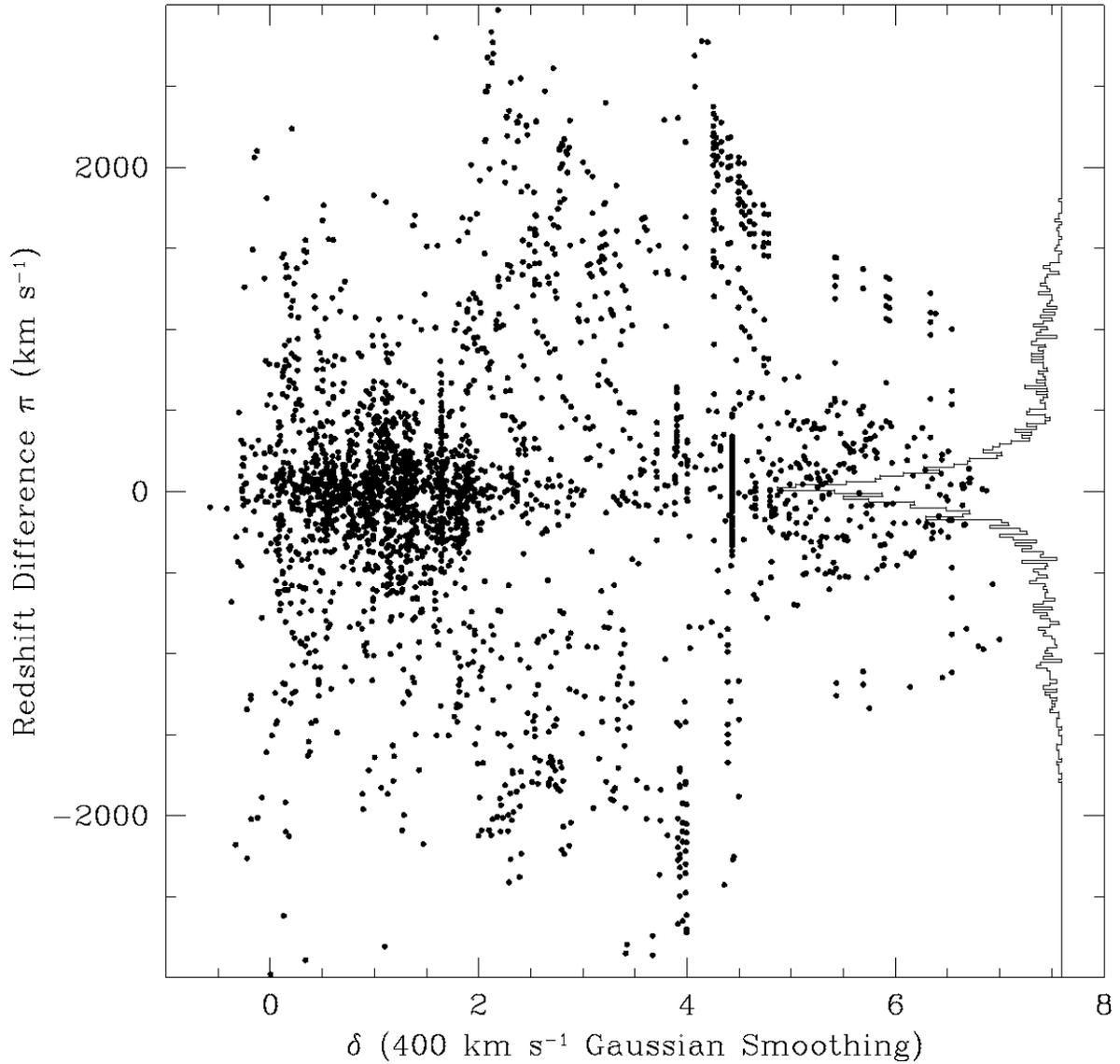}
\caption[]{The observed distribution of differences of redshift of
pairs of galaxies with projected separation less than 1\mpc, from the
volume-limited subsample of the ORS.  Here $\delta$ is the average
overdensity of each pair in units of the sample mean density.  The
marginal distribution of redshift difference is shown as the histogram
on the right-hand side.}  \protect\label{fig:pi-delta}
\end{figure}

Figure~\ref{fig:pi-delta} is a scatter plot of the observed
distribution of redshift differences of galaxy pairs as a function of
local density.  There is substructure in the plot, due to various
discrete structures within the sample volume.  In particular, the
vertical stripes are due to pairs of clusters, all of whose galaxies
are assigned to the same density.  But the overall morphology of this
plot is straightforward to understand. There is a tight core of pairs
whose redshift differences are small (the horizontal concentration of
points along $\pi = 0$), and whose width appears to be an increasing
function of local density, and a very extended background of pairs at
large separation, presumably due to the chance projections, whose
distribution falls off slowly with separation.  The tight core seems
to disappear in the range $2 < \delta < 4$ (we will see this manifest
itself in the following figure), due to the lack of clusters of the
appropriate density in the relatively small volume surveyed.  At very
large densities, the contribution from the background disappears,
because a chance projection between a cluster and field galaxy will
have less than the highest possible density.

\begin{figure}
\plotone{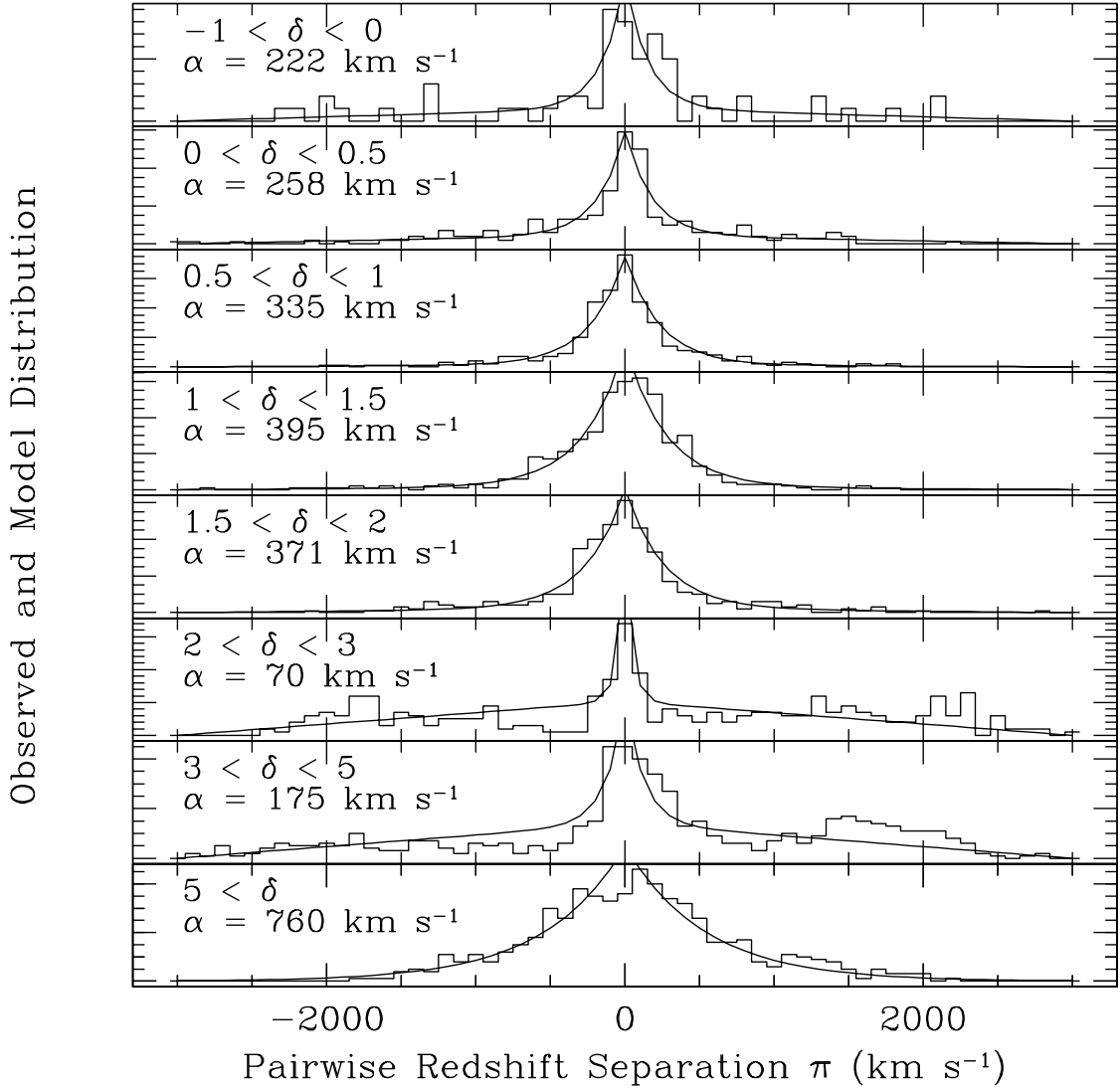}
\caption[]{Histograms of the distribution of redshift differences
$\pi$ as a function of overdensity $\delta$.  The
smooth curves are a maximum likelihood fit of the data in each density
bin to a model of an exponential pairwise velocity distribution plus a
smooth background.}
\protect\label{fig:maxlikefit}
\end{figure}

  Our next step is to quantify the tightness of the central
core. We do this by examining the distribution of values of $\pi$ within
bins of local density $\delta$, as is shown in
Figure~\ref{fig:maxlikefit}.  Within each bin of $\delta$, we make a
maximum likelihood fit of the observed distribution to a model of 
a pairwise velocity distribution plus a smooth background.  Before
describing the model in detail, we note that the fit is not to the
binned data shown here for graphical purposes, but to the individual
galaxy pairs of Figure~\ref{fig:pi-delta}. 

It is argued on both empirical (F94, Marzke \etal\ 1995) and
theoretical (Cen \& Ostriker 1993b; Sheth 1996; Mo, Jing, \& B\"orner
1997; Seto \& Yokoyama 1997; Juszkiewicz, Fisher, \& Szapudi, private
communication) grounds that the distribution function of radial
peculiar velocity differences between close pairs of galaxies should
be exponential.  Of course, the {\em redshift\/} differences of close
pairs include both the effects of peculiar velocities and of true
physical separation (and therefore the redshift difference
distribution function can be modeled as a convolution of the
correlation function with the pairwise distribution function; cf.,
F94), we will ignore this detail here and simply assume that the
redshift difference distribution is exponential as well.  This has the
effect of {\em overestimating\/} the effect of the peculiar velocity
dispersion (although the effect is small, as we will see); we thus
will be able to put upper limits on the peculiar velocity distribution
width as a function of density.

We have found empirically from Monte-Carlo tests that the background
distribution is well-fit by a term proportional to $3000 \kms -|\pi|$,
therefore our model contains only two parameters:
\begin{equation} 
f(\pi) \propto {3000 \kms - |\pi|} + A \exp\left(-{2^{1/2}\,|\pi| \over
\alpha}\right)\quad,
\label{eq:model} 
\end{equation}
where $A$ represents the relative amplitude of the central exponential
relative to the background, and $\alpha$ is the quantity we are
interested in, the second moment of the peculiar velocity
distribution.  The overall normalization of $f$ is determined by the
requirement that it integrate to the observed number of pairs in the
density bin in question.

The results of this fit to each bin are shown as the smooth curves
in Figure~\ref{fig:maxlikefit}; the derived values of $\alpha$ are
given in each panel.   We could calculate formal error bars on
$\alpha$ by finding the values at which the likelihood falls to
$e^{-1/2}$ of its peak value, but the Monte-Carlo experiments we
describe in the next section show this procedure underestimates the
errors 
by factors of three or four.  This is probably due to mismatch of our
simplistic model with the data; in particular, the background
distribution in some bins of 
$\delta$ clearly shows structure which our simple model does not match
(cf., the bin with $3 < \delta < 5$).  However, in every case, our
model does do an adequate job of fitting the shape of the central
peak of the observed distribution function, which is the part in which
we are most interested. 

As expected, the fitted distribution widths are an increasing function
of density, at least to $\delta = 2$.  At the next two bins, the
distribution narrows; there are simply very few regions in the survey
at this density, and in particular, none that are virialized with
substantial velocity dispersion.  This reflects the increasing cosmic
variance as one examines bins of larger $\delta$.  Finally, the width
in the densest bin is very large, over 500 \kms, as we would expect
given that this bin mostly contains galaxies in clusters.  Indeed,
most of the pairs in this bin are found in the Virgo cluster, which
happens to be the richest cluster within the adopted sample volume to
3000 \kms.  This bin contains the largest number of galaxy pairs,
which means that it carries much of the weight in standard measures of
$\sigma_{12}$.  If we fit the model in Eq.~(\ref{eq:model}) to all the
pairs of the sample (i.e., to the full distribution of
Figure~\ref{fig:pi-delta}), we find a central dispersion of $\alpha =
410 \kms$.

At low densities, the distribution of redshift differences is
quite narrow, consistent with the papers quoted above of a
very quiet flow field outside of clusters.  The quantity whose
distribution we are plotting has contributions both from peculiar
velocities, and from the separation of galaxies in the radial
direction in real space. 
We could model this formally as a convolution of the real-space
correlation function with the velocity difference distribution
function, {\em if we knew how to calculate the real space correlation
function for the subset of those galaxies in a given local density
range.}  We will see in the next section that this effect is small;
$\alpha$ as we measure it is in fact a decent approximation to the
true small-scale peculiar velocity dispersion. 

\section{Comparisons with Simulations}
\label{sec:nbody}
  In this section, we use hydrodynamic and $N$-body cosmological
simulations for two purposes.  We first compare the results of our redshift
difference statistic to the right answer, which we know {\it a
priori\/} for the simulations.  After demonstrating that the redshift
difference statistic is indeed measuring something close to the true
small-scale velocity dispersion of galaxies, we can further use the
simulations to compare our observed results with models. 

On the 1-2\mpc\ scales we are considering here, the small-scale
clustering and peculiar velocity properties of galaxies will be
greatly affected by the details of one's model for the relative
distribution of galaxies and dark matter.  In this highly non-linear
regime (in the sense that $\vev{\delta^2} \gg 1$), the simple {\it
ansatz\/} of linear biasing ($\delta_{\rm galaxies} = b\, \delta_{\rm
dark\ matter}$) cannot be expected to hold, nor can we presume that
the velocity fields of galaxies and dark matter trace one another
(velocity bias; cf., Carlberg, Couchman, \& Thomas 1990; Cen \&
Ostriker 1992).  We therefore need simulations that include enough of
the relevant physics to allow us to identify galaxies.  Here we use
the Standard CDM (SCDM) hydrodynamic simulation of Cen \& Ostriker
(1992, 1993ab), which simulated a cube 80\mpc\ on a side with $200^3$
particles and $200^3$ cells.  The parameters of the input power
spectrum are given in Table 1.

We identify galaxies as follows (Cen \& Ostriker 1992): At each time
step, we consider those cells with baryonic overdensity
$(\delta\rho/\rho) > 5.5$ as candidates for regions within which
galaxy formation will occur.  If a cell also satisfies the 
following criteria: the region is collapsing in real coordinates, the
cooling time is less than its dynamical time, and the baryonic mass is
larger than the Jean mass for its density and temperature, the gas
must collapse towards the center 
of the cell with subsequent condensation 
into a stellar system.  So we remove from the gas in the cell in
question the mass that would collapse in timestep $\Delta t$ and create a
collisionless particle at the center of the cell, giving it the same
proper velocity as the gas in the cell.  After creation, these new
particles are treated dynamically as dark matter particles.  The three
components (gas, galaxies and dark matter) interact through gravity.
At any epoch of interest we construct the galaxies out of the
collisionless sub-units formed in cooling collapsing regions using an
adaptive, friends-of-friends linking scheme (Suto, Cen \& Ostriker
1992) to join together neighboring particles into galactic units with
the linking parameters chosen such that the galaxy mass function fits
observations (Cen \& Ostriker 1993b).  

Given this galaxy list, we are ready to make a simulated redshift
survey.  We choose a candidate to
represent the Local Group as we have done in previous papers (e.g.,
Strauss \etal\ 1992, 1993): the galaxy must have a peculiar velocity
in the range 520 to 720 \kms, and have a local density $\delta$
between $-0.2$ and 1.  We then select galaxies, with mass greater than
$10^{9.8}\,M_\odot$, within a sphere of radius 3000 \kms\ around it,
at the number density appropriate for each region in the ORS.  The ORS
excluded zones (especially the zone of avoidance) are put in, and the
effects of extinction on the galaxy number density is also included,
using the Burstein-Heiles (1982) extinction maps.  We calculate the
400 \kms\ Gaussian smoothed density field around each mock galaxy in
redshift space, first collapsing conspicuous clusters as we did in the
real universe.  The resulting galaxy catalog is then put into the
identical code used above to calculate the velocity dispersion
statistic, and the results are tabulated.  One hundred mock realizations of
the ORS sample are generated.
  
\begin{deluxetable}{lrrrrrl}
\tablecaption{Simulations Used}
\tablecolumns{5}
\tablehead{
\colhead{Model} & \colhead{$\Omega_0$} & \colhead{$\Lambda$} &
\colhead{$h$} & \colhead{$\sigma_8$}  & \colhead{Boxsize} & \colhead{Comments}\nl}
\startdata
SCDM & 1.00 & 0.0 & 0.50 & 0.77 & 80\mpc &Hydro simulation, $\Omega_b = 0.06$\nl
LCDM & 0.40 & 0.6 & 0.65 & 0.79 & 128\mpc &$N$-body simulation; Cold
Dark Matter\nl
OCDM & 0.35 & 0.0 & 0.70 & 0.80 & 128\mpc &$N$-body simulation; Cold
Dark Mattter\nl
HDM  & 1.00 & 0.0 & 0.50 & 1.05 & 128\mpc &$N$-body simulation; Hot
Dark Matter\nl
POW1 & 1.00 & 0.0 & 0.50 & 1.05 & 128\mpc &$N$-body simulation; $P(k)
\propto k^{-1}$\nl
POW2 & 1.00 & 0.0 & 0.50 & 1.05 & 128\mpc &$N$-body simulation; $P(k)
\propto k^{-2}$\nl
\enddata

\end{deluxetable} 
{

For the simulation, we of course know the true peculiar velocity of
every galaxy, and therefore we can compare the true difference in
peculiar velocity, $\Delta v_p$ for every pair with projected
separation $< 1\mpc$, to the redshift difference $\pi$.  This
comparison is shown in the upper-left panel of
Figure~\ref{fig:brightgaltests}.  For pairs 
with projected separation this small, the redshift-space difference is
indeed an impressively good measure of the peculiar velocity
difference of galaxies.  The difference $\pi - \Delta v_p$ is shown in
the upper right panel, as a function of local density.  The vast
majority of the points form a tight core with a dispersion less than
100 \kms.  The background is more severe at low densities; chance
projections are more important there, as our intuition would tell us.
This forms the background that we take out in the model of
equation~(\ref{eq:model}).

  For any Monte-Carlo realization of the ORS sample, we can directly
calculate the standard deviation $\sigma$ of peculiar velocity
differences of the close galaxy pairs chosen in a given density bin,
and compare it directly to the value derived from the distribution of
$\pi$.  This comparison is made in the lower-left panel for the 100
mock realizations, at each value of density.  The agreement between the two
quantities is excellent.  The lower-right panel shows the fractional difference
between $\alpha$ and $\sigma$ as a function of local density.  The
derived velocity dispersion is slightly biased upwards (as we would
expect, given the argument at the end of the previous section) by a
mean of 30\% at low densities, and appreciably less at high densities.  The
scatter is impressively small, especially at high densities.

\begin{figure}
\plotone{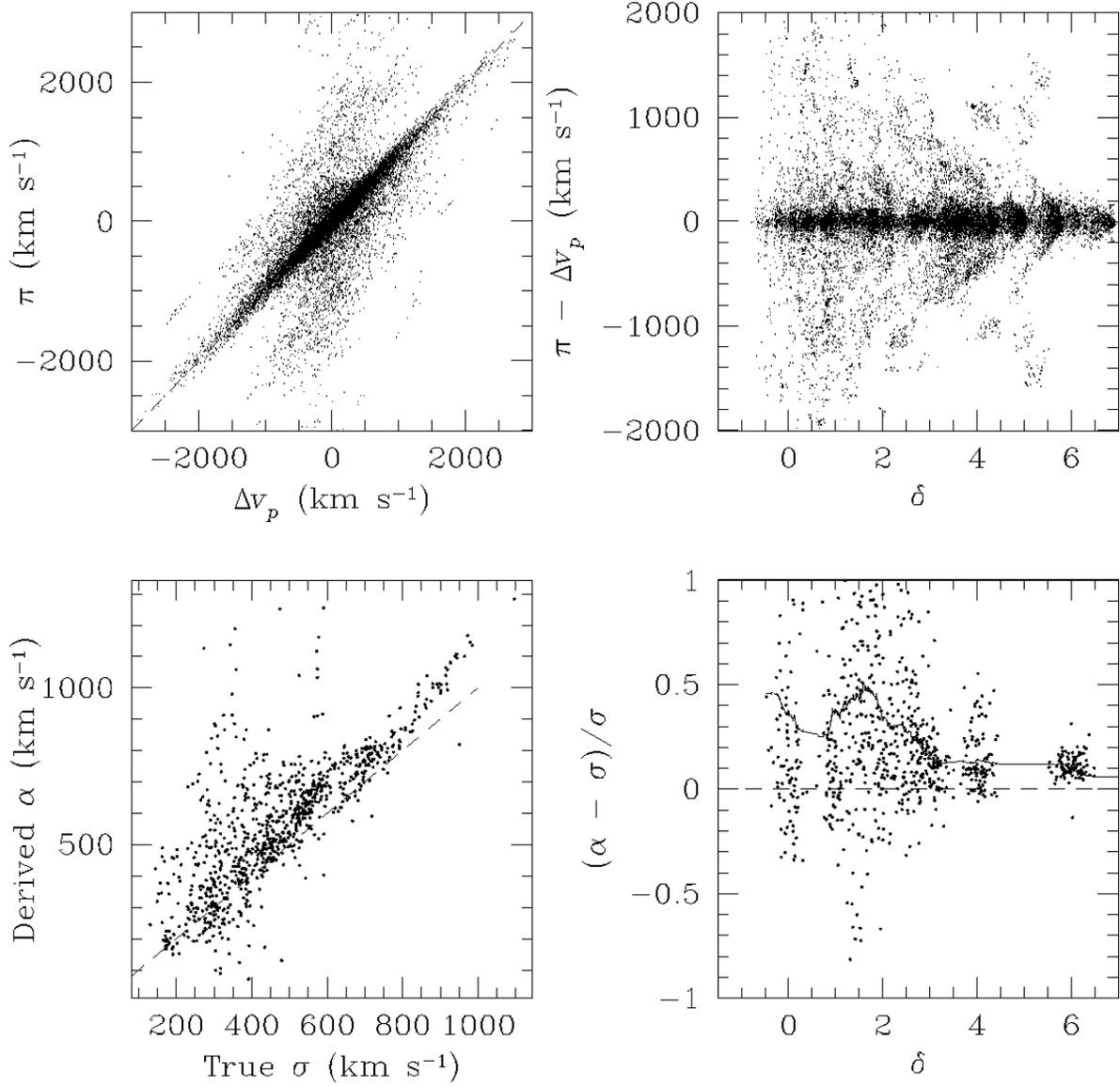}
\caption[]{Comparisons of those statistics measured in redshift space
with the truth, for 100 mock realizations of the ORS sample drawn from a
hydrodynamic simulation of galaxies in a standard CDM universe.  Upper
left panel: the comparison between $\Delta v_p$, the true peculiar
velocity difference of pairs of galaxies with projected separation $<
1\, h^{-1}$Mpc, with their redshift-space difference, $\pi$.  Upper
right panel: the difference $\pi - \Delta v_p$ as a function of local
density, as determined from a 400 \kms\ Gaussian smoothing.  Lower
left panel: the comparison between the scale-length $\alpha$ of the
redshift-space difference of pairs of galaxies, with the standard
deviation $\sigma$ of their true peculiar velocity difference.  For each
realization, results are shown at a variety of local densities.  Lower
right panel: the fractional difference between these two quantities as a function
of local density.  A small scatter has been added to the ordinate in
this plot to make it easier to read. The running mean is given by the
thin solid line.}
\protect\label{fig:brightgaltests}
\end{figure}

  Thus despite the contamination from projected pairs of galaxies, and
the nonuniform sampling of the ORS, the quantity $\alpha$ measured
from the distribution of redshift differences is a good measure of the
small-scale velocity dispersion of galaxies.  With this assurance in
mind, we make a direct comparison between the observations and
simulations in Figure~\ref{fig:cdm-compare}.  The open points are the
observed values of $\alpha$ as a function of local density for the
real ORS sample.  The small points are the results from each of the
100 Monte-Carlo mock realizations of the ORS data, with a small
scatter added to the ordinate to make them easier to distinguish.
Thus the open circles (data) and the small dots (simulation) may be
compared directly.  The mean and standard deviation over the
realizations at each $\delta$ are given by the large solid points with
their errors\footnote{These error bars are typically three times
larger than the formal errors of the model fit.}.  The data lie
roughly one standard deviation below the simulations in almost all
density bins, the exception, interestingly enough, being the highest
density bin.  The formal $\chi^2$ difference between the observations
and the model is 80.5 for 8 degrees of freedom, assuming Gaussian
errors and no covariance between bins; not surprisingly, this is
dominated by the points at $\delta = 3$ and $\delta = 4$.  Formally,
this rules out the SCDM model at the $7.4\, \sigma$ level.  Of course,
the $\chi^2$ statistic does not reflect the fact that the data lies
systematically {\em lower\/} than the model in all density bins.  Thus
the well-established fact that the standard CDM model over-predicts
the observed small-scale velocity dispersion is not just a reflection
of a mismatch at cluster cores; it extends to the lowest density
regions.

\begin{figure}
\plotone{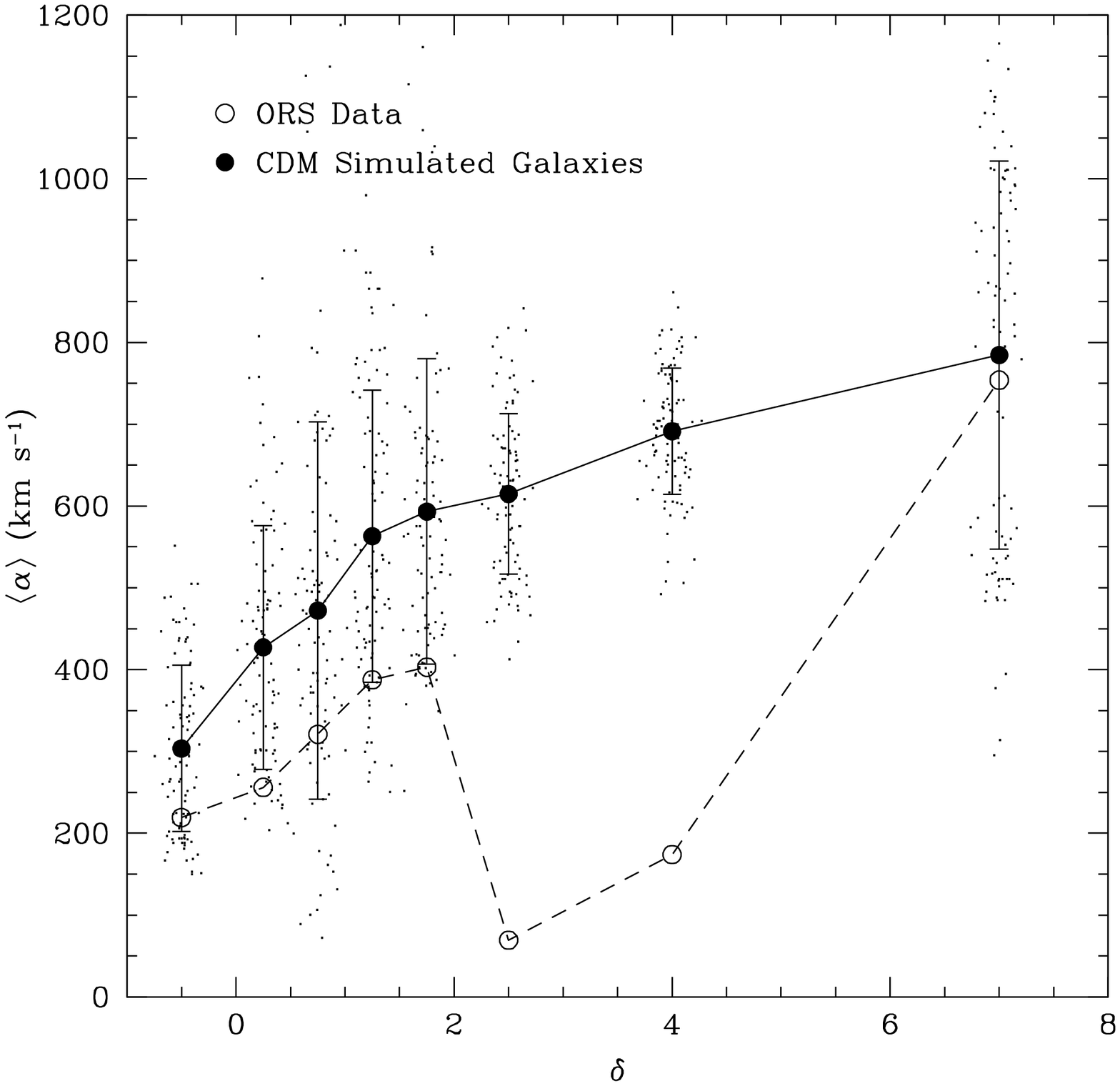}
\caption[]{The open points show the observed velocity dispersion
statistic $\alpha$ as a function of local overdensity, for the real
ORS sample.  The
small points are the results of 100 individual Monte-Carlo
mock realizations of the full ORS sample, drawn from a hydrodynamic
simulation of galaxies in a standard Cold Dark Matter universe. The
solid points with error bars are the mean and standard deviation over
these realizations in each density bin. }
 \protect\label{fig:cdm-compare}
\end{figure}

  What is it about the standard CDM model that causes it to miss the
observed velocity dispersion so dramatically?  The normalization of
the power spectrum in this simulation is roughly 35\% below the COBE
normalization of Bunn \& White (1997) (cf., the discussion in Cen \&
Ostriker 1993b), and was chosen to provide a fit to various observed
phenomena at the several Mpc scale. 
We wish to test the sensitivity of the velocity dispersion statistic
to this normalization.  Rather than repeat the
simulation with a different normalization, we note that the amplitude
grows with time, and so we simply look at this simulation at an
earlier epoch.  Thus, we calculate the
one-dimensional velocity
dispersion $\sigma$ of dark matter particles separated by $< 1\mpc$ from the
CDM simulation at $z = 0$, $z = 0.5$, and $z = 0.7$.  Here we measure
this statistic directly from the simulation, and do not make
Monte-Carlo redshift survey realizations.  For this $\Omega
= 1$ simulation, the linear amplitude of fluctuations is down by
factors of 0.67 and 0.59, respectively, relative to the amplitude at
$z = 0$.  Figure~\ref{fig:galaxies-dark}a shows the velocity
dispersion as a function of local density.  At higher redshifts, the
dynamic range of densities is smaller than that at lower redshifts, of
course, but at a given overdensity, the velocity dispersion is unchanged.
Therefore, the mismatch between the observed and predicted peculiar velocity
dispersions as a function of density in the standard CDM model cannot
be fixed by changing the normalization of the model (cf., the
discussion in Kepner \etal\ 1997).  Note that this is not the case for
the small-scale velocity dispersion averaged over density, as
traditionally defined; indeed, Davis \etal\ (1985) used the observed
small-scale velocity dispersion to set the bias of the Standard CDM
model to $b = 2.5$ (equivalently, $\sigma_8 \approx 0.4$).

  Large-scale hydrodynamic simulations with galaxy formation at the
level of sophistication of that used above are quite time-consuming to
carry out, and we only have the single SCDM model at our disposal for
the comparison with observations.  We argued above that simulations
which include galaxy formation are necessary to compare to
observations, given our lack of knowledge of the nature of biasing on
small scales.  Nevertheless, it is interesting to ask how the velocity
dispersion as a function of density of galaxies is related to that of
dark matter. 
Figure~\ref{fig:galaxies-dark}b compares the
one-dimensional velocity 
dispersion as a function of density for galaxies and dark matter, from
the standard CDM simulation.  There are two physical effects which
cause the two to differ: density bias, which causes changes in the
abscissa of the graph, and velocity bias, which causes changes to the
ordinate.  One can measure these two effects directly from the
simulation itself (Cen \& Ostriker 1992).  At our density smoothing
scale, the density bias is roughly 1.6; that is:
$\delta_{\rm galaxies} = 1.6\, \delta_{\rm dark\ matter}$.  The velocity
dispersion on small scales of galaxies is $\sim 80\%$ that of the dark
matter (velocity bias; cf., the discussion in Carlberg \etal\ 1990;
Cen \& Ostriker 1993b; Carlberg 1994; Summers, Davis, \& Evrard 1995).
Rescaling the two axes of the graph by these factors for the dark
matter gives  Figure~\ref{fig:galaxies-dark}c.  The
two curves are now in qualitative agreement, telling us that at least
to first order, the difference in the velocity dispersion as a
function of local density between galaxies and dark matter in the
simulations can be understood in terms of linear velocity and density
bias. 

\begin{figure}
\plotone{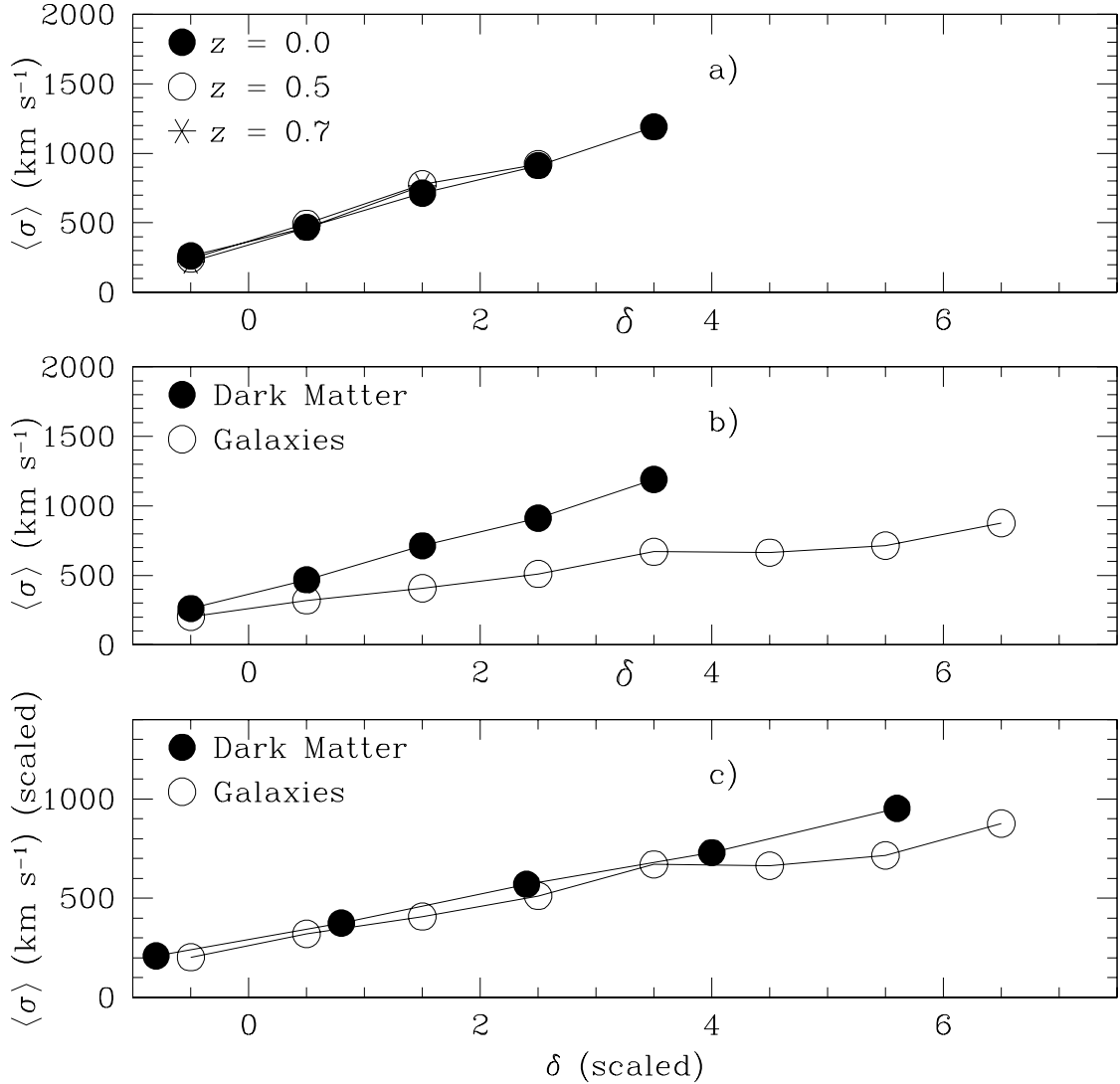}
\caption[]{a. The one-dimensional velocity dispersion on $1\,h^{-1}$Mpc scales as a
function of local density for
$N$-body points from simulations of a standard CDM universe, at three
different redshifts. b.  The dark matter velocity dispersion as a
function of dark matter density at $z = 0$, compared with the velocity
dispersion of galaxies from the same simulation, as a function of the
galaxy density.  c.  As in (b), but with the dark matter density and
velocity dispersion scaled for a density bias $b = 1.6$ and a velocity
bias of 80\%.}
\protect\label{fig:galaxies-dark}
\end{figure}

  We now use dark matter $N$-body simulations of a variety of power
spectra to examine the dependence of the velocity
dispersion-overdensity relation on cosmological model.  These will not
be useful to compare with observations, given our lack of
understanding of the velocity and density bias in each of these
models, but they do show how strongly the velocity dispersion statistic can
distinguish between models in principle.  The dark matter simulations
are done in a 128 \mpc\ box with a $720^3$ PM grid and $240^3$
particles.  The particle mass is $4.2 \times 10^{10} \Omega_0\,h^{-1}
M_\odot$ and the spatial resolution is $0.44 \,h^{-1}$Mpc ($\sim
2.5$ grid cells), both adequate for our purposes.  The details of the
power spectra assumed are given in Table~1.
Figure~\ref{fig:veldisp-nbody} shows the resulting velocity dispersion
as function of overdensity.

Some of the results can be understood intuitively.  The velocity
dispersion of the HDM model in the high-density nonlinear caustics is
very high, but in the low-density regions, no nonlinear structures
have formed, and thus the effective power spectrum on small scales is
very small.  Thus the velocity dispersion becomes vanishingly small at
low densities, in disagreement with what is observed
(Figure~\ref{fig:cdm-compare}).  The two low $\Omega_0$ CDM models
differ only in the presence or absence of a cosmological constant;
this makes essentially no difference to their dynamics today (cf.,
Lahav \etal\ 1991).  With lower $\Omega_0$, the CVT says that these
models have appreciably smaller small-scale velocity dispersion than
does the standard CDM model, bringing them into better agreement with
the observations (compare these curves with that in
Figure~\ref{fig:cdm-compare}).  Finally, the $P(k) \propto k^{-2}$
model has less small-scale power, and therefore a smaller velocity
dispersion, than does the $P(k) \propto k^{-1}$ model.  These models
have a smaller fraction of their volume at low density than do the
others, and thus the velocity dispersion is systematically higher.
Note that as we would expect from the CVT (cf., Kepner \etal\ 1997),
all $\Omega_0 = 1$ models have similar velocity dispersions at high
densities; they are ordered roughly by the value of the rms mass
fluctuations on 1\mpc\ scales in each model.

  Figure~\ref{fig:veldisp-nbody} shows results for dark matter
particles; as discussed above, we need a model for density and
velocity bias of galaxies in each model in order to compare with
observations.  We would expect {\em a priori\/} that the density bias
should be smaller in the low $\Omega_0$ CDM models than in the SCDM
model, from two arguments.  First, as emphasized by Chiu, Ostriker, \&
Strauss (1997), the abundance of rich clusters and comparisons of the
galaxy density and velocity fields constrain the quantity
$\Omega_0^{0.6}/b$; thus the low-$\Omega_0$ models should have lower
bias.  Second, in a low-$\Omega_0$ universe, structure freezes out at
high redshift, and thus the baryons have an extended period in which
to fall into the dark potential wells.  Therefore, at the present, we
expect the baryons to have ``caught up'', and thus show a value of $b$
close to unity (cf., Fry 1996).  Similarly, we would expect the
velocity bias to be smaller as well.  The scaling arguments of
Figure~\ref{fig:galaxies-dark} show that the SCDM, OCDM, and LCDM
{\em galaxy\/} velocity dispersion as function of {\em galaxy\/} density may thus be
more similar to one another than Figure~\ref{fig:veldisp-nbody} would imply.  It
would be amusing indeed if they become degenerate, but it will require
full galaxy formation simulations of the low-$\Omega_0$ models to
determine if this is indeed the case. 

\begin{figure}
\plotone{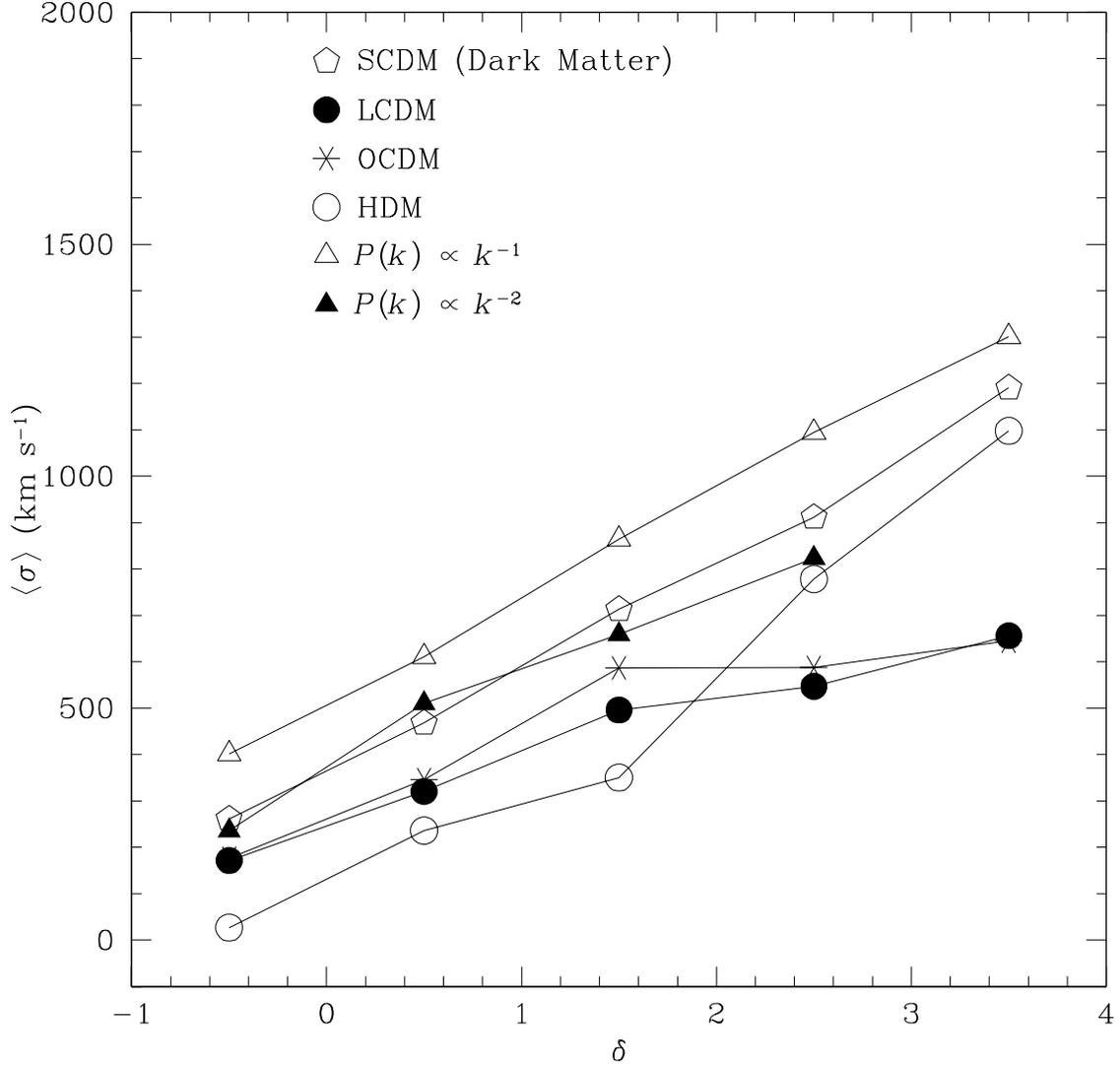}
\caption[]{The one-dimensional pairwise velocity dispersion on $1\,h^{-1}$Mpc
scales as a function of local density for $N$-body points from
simulations of a series of different cosmological models.  The SCDM
points are from dark matter points in the SCDM hydrodynamic
simulation.}  \protect\label{fig:veldisp-nbody}
\end{figure}

\section{Discussion and Conclusions}
\label{sec:conclusions}

We have developed and tested a method to measure the small-scale
velocity dispersion of galaxies from a redshift survey, as a function
of local density.  This approach is much less sensitive to the
presence or absence of a rich cluster in the survey volume than is the
traditional velocity dispersion of galaxies.  As expected, the
observed velocity dispersion of galaxies is an increasing function of
local density, varying between 220 \kms\ at low densities, to 760
\kms\ at high densities.  The small value of the velocity dispersion
at low densities is consistent with reports of a cold velocity flow of
galaxies outside of clusters.  For example, Willick \etal\ (1997)
report from an analysis of Tully-Fisher data that the pairwise velocity
dispersion of spiral galaxies of $175 \pm 30 \kms$, in
good agreement with our value at low densities.  Davis \etal\ (1997)
find a pairwise velocity dispersion of \iras\ galaxies of $140 \pm 22
\kms$ in their single-particle-weighting technique, which is well {\em
below\/} our value; their value for optically selected galaxies, $180
\pm 22\kms$, is in good agreement with ours. 

We have found that the long-standing discrepancy between observed
small-scale velocity dispersion of galaxies, and the predictions of
the standard CDM model with COBE-normalization, is apparent even when
plotted as a function of local density.  Thus the problem cannot be
``solved'' by restricting attention to any particular regime of
overdensity. 

This did not have to be the case; it is well-known that the abundance
of rich clusters in standard CDM is appreciably larger than the
observed value (e.g., Bahcall \& Cen 1992); this is closely related to
the fact that the observed abundance of clusters puts a tight
constraint on the combination $\sigma_8 \Omega^{0.56}$ (cf., White,
Efstathiou, \& Frenk 1993; Eke, Cole \& Frenk 1996; Viana \& Liddle
1996; Pen 1996; Fan, Bahcall, \& Cen 1997; Chiu \etal\ 
1997), giving a value about half that appropriate for COBE-normalized
standard CDM (a problem that {\em can\/} be addressed by altering the
normalization).  As we argued in the introduction, the standard velocity
dispersion statistic is heavily weighted by clusters, meaning that if
a model overpredicts the distribution of clusters, it is bound to
overpredict the galaxy velocity dispersion.  Plotting the velocity
dispersion as a function of local density removes this problem, and
yet the standard CDM model still overpredicts the observed velocity
dispersion, even at low densities.  Our Monte-Carlo experiments show
that in any single bin of density, the observations do not rule out
SCDM at an interesting level, but the velocity dispersion is
overpredicted consistently by the model in all bins.  If we treat the
distribution of values in the simulations as Gaussian, and the
different bins as independent, we formally rule out CDM at the $7.4
\,\sigma$ significance level. 

 The predicted velocity dispersion as a function of density is quite
insensitive to the normalization of the power spectrum.  The
small-scale velocity dispersion as a function of local density is
quite different when calculated from dark matter particles and
galaxies in a hydrodynamic simulation; this may be understood to first
order as due to a combination of density and velocity bias.  We need
sophisticated galaxy-forming hydrodynamic simulations of a series of
models in order to make further comparisons of observations with
models.  However, rough comparisons with $N$-body simulations suggests
that the HDM model cannot match the observed velocity dispersion,
while low-$\Omega_0$ CDM models should fare better than standard CDM.
We are in the process of running hydrodynamic simulations with roughly
three times the spatial resolution, nine times the mass resolution,
and improved cooling models for a variant of the LCDM model;
%Omega=0.37, Lambda=0.63, H=70, Omega_b=0.049, sigma_8=0.79;
we look forward to measuring the velocity dispersion as a
function of overdensity for this simulation, and comparing with
observations. 

  The ORS is the best available redshift survey for calculating the
velocity dispersion statistic.  The \iras\ 1.2 Jy survey (Fisher
\etal\ 1995) is too sparse to give an interesting number of pairs at
small separation.  It is possible that the \iras\ 0.6 Jy survey (the
PSC-z; cf, Efstathiou 1997) will be well-suited for our statistic.  In
the future, we could imagine using a volume-limited subsample from the
Sloan Digital Sky Survey (cf., Gunn \& Weinberg 1995; Strauss 1997)
3000 \kms\ thick at $\sim 30,000 \kms$ from the Local Group,
containing of the order of 60,000 galaxies.  This would allow us to
measure this statistic to much higher accuracy over a substantially
larger volume.

\acknowledgements 
  M.A.S. acknowledges the support of the Alfred P. Sloan Foundation,
NASA Theory Grant NAG5-2882, NSF Grant AST96-16901 and the
hospitality of the Astronomy Department of the University of
T\=oky\=o, where much of this work was carried out.  J.P.O. and
R.C. were supported by NASA grant NAG5-2759 and NSF Grant
ASC93-018185.  We thank Ofer
Lahav, Bas\'\i lio Santiago, Alan Dressler, Marc Davis, and John Huchra
(the ORS team) for permission to use the ORS data before publication.

\end{document}